\documentclass[11pt]{article}
\usepackage[utf8]{inputenc}
\usepackage[numbers, square]{natbib} 
\usepackage{amsmath,amssymb}
\usepackage{geometry}
\usepackage[colorlinks=true, allcolors=blue]{hyperref}
\usepackage{bm}
\usepackage{enumitem}
\usepackage{graphicx}
\usepackage{booktabs}
\usepackage{multirow}
\usepackage{tcolorbox}

\title{%
  On the Genealogy of Machine Learning Weather Prediction \\[0.4em]
  \normalfont\normalsize\itshape
  Physics Inherited, Data Forgotten --- Toward a Principled Trade-off in Surrogate Modeling
}

\author{
  Mohammad H. Erfani\thanks{\texttt{moerfani@ucsd.edu}} \\[0.3em]
  \small Center for Western Weather and Water Extremes (CW3E) \\
  \small Scripps Institution of Oceanography \\
  \small UC San Diego
}
\date{}

\begin{document}
\maketitle

\begin{abstract}
Modern machine-learning weather prediction (MLWP) has largely inherited the initial-value-problem (IVP) framing of numerical weather prediction (NWP). This inheritance leads to a dominant paradigm of learned autoregressive time-stepping and constrains how the learning problem is defined and architectures are favored. In this study we make the inheritance explicit, contrast two philosophical traditions: ``scientific surrogate modeling,'' where machine learning (ML) is embedded within a physical system and must respect its structural constraints, and ``free-form data-driven modeling,'' where atmospheric fields are treated as spatiotemporal sequences and models learn latent dynamics without explicit physical constraints. By reviewing the governing primitive equations, surveying recent literature, and analyzing concrete physical examples, we map each modeling paradigm to either a state-conditioned or evolution operator formulation. We conclude that principled model selection requires explicitly aligning architecture and training objectives with either the physical system structure or the statistical structure of the data.
\end{abstract}

\section{Introduction}
Traditionally, physics has been responsible for explaining the Earth as a dynamical system. Fundamental physical laws such as Newton's Second Law of Motion and the First Law of Thermodynamics govern the evolution of the atmosphere. These laws are converted into mathematical equations that form the foundation of NWP~\citep{kalnay2003atmospheric, lynch2008origins}.

Because physics has historically dominated weather prediction, MLWP inherited not only atmospheric datasets and forecasting targets from NWP, but also many of its conceptual assumptions~\citep{dueben2018challenges}. In particular, modern MLWP adopt the ``scientific formulation'' of weather prediction as an IVP~\citep{scher2018toward, weyn2019can}. Consequently, many MLWPs are designed as learned time-stepping systems that autoregressively evolve atmospheric states forward in time~\citep{keisler2022forecasting, pathak2022fourcastnet, bi2023accurate, lam2023learning, chen2023fuxi, lang2024aifs}. This inheritance is not necessarily misguided. Weather prediction is fundamentally temporal, and any successful MLWP system must respect the time-evolving nature of the atmosphere. However, the inherited framing also constrains the way the ``learning problem'' itself is defined. Instead of first asking what the structure of the data suggests from a ML perspective, MLWP often begins by reproducing the essence of numerical solvers: prognostic variables, diagnostic variables, tendencies, and time-marching operators. 

Conversely, if the statistical structure of atmospheric data were allowed to dictate the problem formulation, the resulting models would look quite different. From a purely ML perspective, weather forecasting is fundamentally a spatiotemporal sequence prediction task, i.e., conceptually akin to video frame extrapolation~\citep{shi2015convolutional, wang2017predrnn, wang2018predrnn++, wang2019memory, yu2020efficient, guen2020disentangling, lin2020self, gao2022simvp, tan2023openstl}. To capture the profound spatial and temporal dependencies inherent in the data, a purely data-driven approach favors end-to-end trainable architectures that seamlessly integrate spatial and temporal modules, unconstrained by the need to emulate physical time-marching operators. This is investigated by presenting the atmospheric data without their physical semantics to different AI chatbots and asking which models they would recommend (see Appendix~\ref{sec:appendix-a}).

Consequently, depending on the foundational philosophy adopted, the development of modern MLWP can be seen as bifurcating into two distinct traditions:

\begin{enumerate}
\item Scientific surrogate modeling, where ML is embedded inside an existing physical system and must preserve its structure.
\item Free-form data-driven modeling, where the task is defined primarily by the statistical structure of the data itself.
\end{enumerate}

Before proceeding further, it is important to distinguish between two fundamentally different use cases of NWP: weather prediction and climate projection. From a physical standpoint, weather and climate are worlds apart~\citep{watson2021machine}. A commonly cited distinction is that weather prediction is framed as an IVP, whereas climate projection is framed as a boundary value problem (BVP)~\citep{nguyen2023climax}. Climate models, more precisely referred to as Earth System Models (ESMs), not only include the atmospheric model but couple the cryosphere, land surface, and ocean components, to represent the gold standard of long-range climate modeling. From a mathematical standpoint, however, both framings share a common skeleton: the primitive equations of atmospheric motion, which constitute a system of partial differential equations (PDEs). As with any PDE, a unique solution requires auxiliary information to close the system. This information is supplied with respect to the independent variables: either with respect to time (in the form of initial conditions specifying the full system state at $t = 0$) or with respect to space (in the form of boundary conditions constraining the solution at the edges of the domain). Consequently, the governing PDE structure is the same, and a ML surrogate operates on that structure regardless of whether the downstream application is weather forecasting or climate projection.


This shared mathematical structure is precisely what modern MLWP has inherited. It relies on autoregressive rollout paradigm and IVP framing of weather prediction. This study is supposed to raise an important question: is MLWP best understood as a scientific surrogate that must preserve the structure of the physical system, or as a free-form data-driven learner guided by the statistical structure of the data? These two philosophies lead to very different modeling choices. If atmospheric observations were presented without any physical interpretation, as high-dimensional spatiotemporal sequences alone, the ML solution would likely look very different from the approaches that have become standard in the field. The purpose of this work is to make this inheritance explicit, examine both traditions critically, and ask which approach better surrogates the underlying physical model in terms of stability and fidelity.

To answer this question, we examine the existing literature and revisit the mathematical equations governing the atmosphere and its physical processes. We structure our analysis as follows. We first survey the literature and the progress made since MLWP emerged as a field. We then review NWP to understand how its governing equations are numerically integrated. Next, we define various data-driven operators and, building on our NWP review, categorize different ML approaches according to how they treat prognostic versus diagnostic variables. Finally, we trace the origins of today's ML weather models to understand why they behave as learned NWP time-steppers rather than generic spatiotemporal sequence models. By following this history, we show how the inherited IVP framing has shaped current practice and highlight the need for a more balanced approach that also embraces data-centric modeling.

\section{Literature Review}
Starting from 2018 and 2019~\citep{dueben2018challenges, scher2018toward, weyn2019can}, modern MLWP inherited its dominant autoregressive rollout paradigm from the historical framing of NWP as an IVP. Recent studies have continued to follow traditional numerical solver approaches while working to better represent global grid data. These efforts have explored various representations including equiangular gnomonic cubed spheres~\citep{weyn2020improving}, Fourier neural operators~\citep{pathak2022fourcastnet}, spherical harmonic functions~\citep{bonev2023spherical}, multi-mesh graph structures~\citep{lam2023learning}, and customized, geometry-aware tokenizers~\citep{bi2023accurate, nguyen2023climax}. The primary aim of these approaches is to better capture spatial relationships across three-dimensional atmospheric grids by incorporating ML modules with stronger inductive biases for grid-point connectivity. Despite these architectural advances, however, the temporal dimension has received comparatively little attention. This approach represents a major limitation for stable long-term forecasting, particularly for atmospheric dynamics, which are fundamentally time-dependent processes. Some studies address this problem in the same autoregressive framework by introducing architectural and training-level interventions that constrain the temporal evolution of the predicted states~\citep{watt2023ace, guan2025lucie}. 

In parallel, spatiotemporal modeling has also gained attention across different components of Earth system modeling~\citep{yu2024deep}. In climate science, to estimate annual mean global distributions of temperature given a wide range of emissions (e.g., carbon dioxide, methane, and aerosols) the importance of spatial and temporal dependencies was recognized through a model that sequentially processes spatial data before feeding it into a temporal module~\citep{watson2022climatebench}. This approach was further refined and better aligned with the common practice of spatiotemporal modeling~\citep{nguyen2023climax}. Such an approach, i.e., predicting the target variable at timestep $t$ given the preceding sequence of driving forces, was initially introduced in hydrology for soil moisture and rainfall–runoff modeling in lumped systems using LSTM~\citep{fang2017prolongation, kratzert2018rainfall}, and expanded to atmospheric tracer transport using a spatiotemporal model~\citep{erfani2025interactive, erfani2026spatiotemporal}. 

Overall, it appears that experimental setups in which the model predicts a target sequence given a preceding input sequence (formally, $\mathcal{F} : \mathbf{X}^{T} \mapsto \mathbf{Y}^{T'}$, where $T$ denotes 
an arbitrary-length sequence of past time steps used to approximate a future sequence of the target variable, $T'$) are conventionally handled by spatiotemporal models. In contrast, setups in which the model forecasts the next state of a variable given its current state (formally, $\mathcal{F} : \mathbf{X}_t \mapsto \mathbf{X}_{t + \Delta t}$) align with the autoregressive rollout paradigm inherited from the IVP framing of NWP. In the following sections, we first examine how each of these mathematical mappings is adopted in the physical components of NWP, and then explore their corresponding learned operators in the context of ML.

\section{Numerical Weather Prediction} \label{sec:nwp}
The following overview is intentionally schematic. Its purpose is not to reproduce the full complexity of operational NWP, but to isolate the parts of the system that matter for the surrogate-modeling argument. Vilhelm Bjerknes first recognized that NWP was possible in principle in 1904. He proposed that weather prediction could be viewed as an IVP in mathematics: since physical laws govern how meteorological variables evolve over time, if we possess an accurate representation of the atmosphere’s initial state, we can numerically integrate these governing equations forward in time to generate a forecast~\citep{meted2020impact}. 

\subsection{The Primitive Equations}
At its core, NWP involves solving a set of partial differential equations commonly referred to as the \textbf{Primitive Equations}. These equations are designed to resolve six fundamental variables: three-dimensional wind velocity components ($u, v, \omega$), temperature ($T$), moisture ($q$), and geopotential height ($z$).

\subsubsection*{Wind Forecast Equations}
\begin{equation}
\frac{\partial u}{\partial t} = - u \frac{\partial u}{\partial x} - v \frac{\partial u}{\partial y} - \omega \frac{\partial u}{\partial p} + f v - g \frac{\partial z}{\partial x} + F_x
\label{eq:wind_field_u}
\end{equation}

\begin{equation}
\frac{\partial v}{\partial t} = - u \frac{\partial v}{\partial t} - v \frac{\partial v}{\partial y} - \omega \frac{\partial v}{\partial p} - f u - g \frac{\partial z}{\partial y} + F_y
\label{eq:wind_field_v}
\end{equation}

\subsubsection*{Continuity Equation}
\begin{equation}
\frac{\partial u}{\partial x} + \frac{\partial v}{\partial y} + \frac{\partial \omega}{\partial p} = 0
\label{eq:continuity}
\end{equation}

\subsubsection*{Temperature Forecast Equation}
\begin{equation}
\frac{\partial T}{\partial t} = - u \frac{\partial T}{\partial x} - v \frac{\partial T}{\partial y} - \omega \left( \frac{\partial T}{\partial p} - \frac{R T}{c_p p} \right) + \frac{H}{c_p}
\label{eq:temperature}
\end{equation}

\subsubsection*{Moisture Forecast Equation}
\begin{equation}
\frac{\partial q}{\partial t} = - u \frac{\partial q}{\partial x} - v \frac{\partial q}{\partial y} - \omega \frac{\partial q}{\partial p} + E - C
\label{eq:moisture}
\end{equation}

\subsubsection*{Hydrostatic Equation}
\begin{equation}
\frac{\partial z}{\partial p} = - \frac{R T}{p g}
\label{eq:hydrostatic}
\end{equation}

Because the non-linear prognostic equations do not possess closed-form analytical solutions, we must rely on numerical schemes to solve them. In practice, solving these equations is a process of discrete integration over time and space. We can represent the essence of this integration using a simple Euler forward scheme. If $\psi$ represents any of our \emph{prognostic} variables, its state at time $t + \Delta t$ can be approximated by the current state and its calculated time tendency:

\begin{equation}
\psi(t + \Delta t) \approx \psi(t) + \left( \frac{\partial \psi}{\partial t} \right) \Delta t
\end{equation}

In this framework, the numerical model acts as an engine that calculates the tendency term ($\partial\psi/\partial t$) using the physical laws shown above, then iteratively updates the state of the atmosphere. This time-stepping structure has profoundly influenced the dominant formulation of modern MLWP. A learned model of the form

\begin{equation}
X_{t+\Delta t} = \mathcal{M}_{\theta}(X_t)
\label{eq:mlwp}
\end{equation}

\noindent
behaves analogously to a learned numerical integrator: it repeatedly applies a transition operator to evolve the atmospheric state forward in time.

A closer look at these equations reveals a critical distinction in how different meteorological variables are treated. Equations (\ref{eq:wind_field_u}), (\ref{eq:wind_field_v}), (\ref{eq:temperature}), and (\ref{eq:moisture}) feature a time derivative ($\partial/\partial t$) on the left-hand side. These are known as \textbf{prognostic equations}, where the future states of $u, v, T$, and $q$ are determined explicitly by calculating their dynamic changes over time. Conversely, Equations (\ref{eq:continuity}) and (\ref{eq:hydrostatic}) lack a time derivative. These are \textbf{diagnostic equations}. Variables such as vertical velocity ($\omega$) and geopotential height ($z$) are not evolved forward in time directly. Instead, they are instantaneously ``diagnosed'' at each time step based entirely on the concurrent state of the prognostic variables.

\subsection{Physical Processes}
In the set of NWP equations, some variables, specifically $F_x, F_y, H, E,$ and $C$, represent physical processes that impact our primary variables. These processes are inherently complex; they often involve scales far smaller than the grid spacing of the model (such as individual convective clouds) or rely on physical mechanisms (like radiation transfer) that are too computationally expensive to resolve from first principles. Because we cannot calculate these effects directly within the core equations, we must estimate them using empirical approximations. In numerical modeling, this technical estimation process is known as \textbf{parameterization}. The accuracy of an NWP forecast is fundamentally linked to how well these parameterizations mimic reality.

For surrogate modeling, parameterizations are especially notable because they already reside at the intersection of physics and empiricism. They do not represent independent atmospheric states; instead, they are evaluated directly from the current model state using empirical constants, closure assumptions, and internal sub-grid variables. Crucially, this creates an interactive feedback loop within the model. The dynamical core computes the resolved variables, which are then passed to various parameterization schemes to estimate the unresolved physical processes. The outputs of these schemes are subsequently fed back into the tendency equations as forcing terms, updating and refining the resolved variables for the next time step.

In this sense, \emph{the outputs of parameterization schemes can be viewed as dependent variables that are functions of the resolved fields produced by the dynamical core.} Recognizing parameterizations as functions of the resolved state clarifies the relationship between variables at each time step of the iterative solver. In other words, the prognostic fields that the model directly resolves act as predictor variables (features), while the tendencies and diagnostic closure terms produced by the parameterizations act as predictands (targets). This observation highlights the fundamental difference between prognostic and diagnostic variables which is consistent with the distinction drawn in the literature review. On the ML side, this distinction has direct practical implications in determinding both the learning objective and the appropriate model architecture.

\section{Machine Learning}
Choosing the right ML model requires a systematic approach that balances problem type, data characteristics, and practical constraints such as stability and computational cost.

\subsection{Problem Characteristics}
Given what has been discussed in sections~\ref{sec:nwp}, MLWP, whether at the dynamical core level or for specific physical processes, is generally framed as a \textbf{supervised learning} and \textbf{regression} problem. In this context, the term \emph{prediction} is used broadly in the ML literature to refer to any model output, which can obscure an important distinction relevant to MLWP. This article therefore distinguishes between:

\begin{enumerate}
  \item State-conditioned operator evaluation
  \item Evolution operator learning
\end{enumerate}

\subsubsection*{State-Conditioned Operator Evaluation}
The first formulation is the simpler of the two, a standard regression mapping:

\begin{equation}
Y_i = \mathcal{G}_{\theta}(X_i).
\label{eq:standard-regression}
\end{equation}

At its core, this is a \textbf{functional mapping from inputs to outputs at a single time level}, i.e., a direct, memoryless transformation with no temporal dependencies encoded in the model itself. Each data point is treated as an \textbf{independent observation} within a feature space, making this fundamentally an interpolation problem. The model learns \emph{what the output should look like given the input}. The goal, then, is to approximate the operator $\mathcal{G}_{\theta}$ as faithfully as possible across the input domain. This operator is generally used in the parameterization of physical processes, which is discussed in Section~\ref{sec:parameterization-example}.

\subsubsection*{Evolution Operator Learning}
Evolution operator learning concerns estimating the future state of a dynamical system given its current state. This can take the form of a transition operator,

\begin{equation}
X_{t+\Delta t} = \mathcal{M}_{\theta}(X_t),
\label{eq:autoregressive-formula}
\end{equation}

or a tendency operator,

\begin{equation}
X_{t+\Delta t} = X_t + \Delta t\,\mathcal{F}_{\theta}(X_t),\qquad \frac{dX}{dt} = \mathcal{F}_{\theta}(X_t)
\label{eq:autoregressive-tendency-formula}
\end{equation}

Both formulations can be deployed autoregressively, since the output updates the evolving state. In other words, the model approximates either the transition or the tendency operator of a dynamical system. Unlike state-conditioned operator evaluation, which is fully cross-sectional with respect to the time domain, evolution operator learning is fundamentally sequential, i.e., predictions are conditioned on the historical trajectory of the target variable, making temporal memory an intrinsic component of the formulation. In this strictly autoregressive form, the state evolves solely based on its own past values, independent of external forcings. In broader literature, this is often referred to as ``time-series forecasting'' where the future trajectory is projected entirely via temporal correlations. As discussed in Section~\ref{sec:nwp}, modern MLWP is predominantly formulated using this operator (Equations~\ref{eq:mlwp} and~\ref{eq:autoregressive-formula} represent exactly same formula) but it also applies to other areas discussed in Section~\ref{sec:parameterization-example}.

Modeling sequence evolution does not mean the system must operate in isolation from its external physical environment. In \textbf{multivariate approaches}, external physical drivers (exogenous variables) are explicitly incorporated into the framework:

\begin{equation}
y_t = \beta_1 {x_1}_t + \beta_2 {x_2}_t + \dots + \beta_n {x_n}_t + \phi\, y_{t-1} + \epsilon_t
\end{equation}

\noindent
this allows the model to respond to external physical forcings $({x_1}_t, {x_2}_t, \dots, {x_n}_t)$ while the autoregressive term $\phi\, y_{t-1}$ accounts for temporal memory and delayed system response. In ML, this formulation can be generalized through a parametric state-update equation (Recurrent Neural Networks):

\begin{equation}
h_t = \mathcal{R}_{\theta}(X_t,\, h_{t-1})
\label{eq:rnn-formula}
\end{equation}

\noindent
where $X_t$ represents all driver variables at the current time step, and $h_{t-1}$ encodes whatever information is inherited from previous time steps, typically in a latent space (hidden state) that can ``recurrently'' accumulate context over a sequence length extending well beyond a single lag. The output $y_t$ is then computed from the updated hidden state. This approach represents architectural components specifically designed to capture trends, seasonal patterns, and cyclical behaviors significantly improve the model's ability to make accurate forecasts over extended time periods~\citep{hochreiter1997long, chung2014empirical, van2016wavenet, bai2018empirical, vaswani2017attention} which has been expanded to spatiotemporal modeling as well~\citep{tran2015learning, shi2015convolutional, bertasius2021timesformer, arnab2021vivit}.

\subsection{Data Characteristics}
In data-driven modeling, the data ultimately constrain the choice of model. Understanding data characteristics, therefore, is not a preliminary formality; it is the foundation of the entire modeling process. For MLWP, the defining characteristic is spatiotemporal dependency: the data carry structure in both the time and space domains simultaneously. Recognizing this dependency allows us to translate a physical problem into a well-posed ML task.

With both the problem type and the data structure in hand, one might expect that configuring the experimental setup would be an easy task. However, it is not. The reason involves a more subtle question: who defines the task? The answer shapes not just the model architecture, but the entire framing of the problem. If the task is defined through the lens of physical science, the data are no longer mere tensors; they are atmospheric states, diagnostic variables, tendencies, closures, and physical constraints, each carrying domain-specific meaning. If, instead, the task is framed in a free-form, data-driven context, those same objects are reduced to multichannel spatiotemporal arrays—effectively, a video sequence. These two descriptions are not equivalent in practice: they activate different modeling instincts, favor different architectures, and carry different assumptions about what the model is expected to learn.

Under the scientific formulation, current MLWP often learns a one-step transition operator and rolls it forward using the evolution operator learning framework (Equation~\ref{eq:autoregressive-formula} or~\ref{eq:autoregressive-tendency-formula}). As discussed before, this mathematical formalization mirrors NWP time-stepping~\emph{(Ironically, while this formulation explicitly mimics the structure of physical systems, the resulting ML architectures rarely exhibit any difference in how they treat prognostic versus diagnostic fields. Instead of diagnosing the latter from the former, they typically just advance all variables simultaneously)}.

Under a free-form spatiotemporal formulation, however, one may instead define the problem as sequence-to-sequence prediction:

\begin{equation}
    \mathcal{F} : \mathbf{X}^{T} \mapsto \mathbf{X}^{T'}, \qquad
    \{X_{t-k}, \ldots, X_t\} \mapsto \{X_{t+\Delta t}, \ldots, X_{t+m\Delta t}\}
    \label{eq:seq2seq}
\end{equation}

\noindent
This formalization can be implemented through recurrent models (Equation~\ref{eq:rnn-formula}). While it may still generate future frames autoregressively, it treats the problem as latent spatiotemporal pattern evolution rather than as an explicit learned numerical time-stepper.

\section{Operator Types in Practice: Physical Examples} \label{sec:parameterization-example}
Not all variables in a physical model evolve through time in the same way. As established earlier, prognostic variables have explicit time derivatives governing their evolution. They are marched forward step by step, making their estimation naturally suited to the evolution operator learning framework discussed above. Other variables, however, do not have their own evolution equations. Instead, they are determined entirely by the current state of the prognostic variables at any given time step. These include diagnostic variables, physical tendencies, and the closure terms produced by parameterization schemes. In this section, we investigate different parameterization schemes and discuss which data-driven operators are best suited to surrogate their physical or empirical formulations.  

\subsection{Friction in the Shallow Water Equations}
The one-dimensional \textbf{shallow water equations (SWE)} describe the conservation of mass and momentum for a depth-averaged free-surface flow~\citep{chaudhry2008open}:
\begin{equation}
\frac{\partial h}{\partial t} + \frac{\partial (h u)}{\partial x} = 0
\end{equation}

\begin{equation}
\frac{\partial (h u)}{\partial t} + \frac{\partial}{\partial x}\left(h u^2 + \tfrac{1}{2} g h^2\right) = - g h \frac{\partial z_b}{\partial x} - S_f
\end{equation}

\noindent
where $h$ is water depth, $u$ is depth-averaged velocity, $g$ is gravitational acceleration, $z_b$ is bed elevation, and $S_f$ is the friction slope—a source term encoding energy losses due to bed resistance. The prognostic variables here are $h$ and $u$: they carry explicit time derivatives and are stepped forward in time. Friction, by contrast, represents a physical process that must be parameterized rather than directly resolved. The frictional force arises from shear stress between the flowing water and the channel bed, and it is expressed through the friction slope $S_f$ as:

\begin{equation}
S_f = \frac{C\,V\,|V|^{m-1}}{R^p}
\end{equation}

\noindent
where $C$ and $p$ are coefficients determined by the chosen friction law, $V$ is the flow velocity magnitude, $R$ is the hydraulic radius, and $m$ depends on the flow regime. 

Unlike the prognostic variables  $(h, u)$, $S_f$ has no time derivative of its own. It is computed instantaneously from the current flow state $(h_t, u_t)$ at each time step. Because its value is fully determined by the present prognostic state, it has no temporal memory of its own past values. As a result, learning a surrogate for $S_f$ is a state-conditioned operator evaluation task (Equation~\ref{eq:standard-regression}). The model simply learns to map the current flow conditions to the corresponding instantaneous frictional response. An important clarification here: diagnostic variables do exhibit temporal correlation in observed data, but that correlation is inherited from the prognostic state driving them, not from any autonomous dynamics of their own.




\subsection{Precipitation in the Moisture Forecast Equation}
Precipitation appears in the moisture forecast equation (Equation~\ref{eq:moisture}) as one of the physical process terms. Precipitation is inherently a flux quantity. In many operational MLWPs~\citep{nipen2026regional, bano2025regional}, precipitation is treated as a state-conditioned operator evaluation task (Equation~\ref{eq:standard-regression}) — described as a feature that ``\emph{is only part of the forecast state but is not used as input to the model},'' meaning the model predicts it at each timestep but never ingests it as a feedback input. This treatment is physically justifiable because precipitation functions as a sink term: it is the end product of the resolved variables such as atmospheric dynamics and moisture processes. As shown in Equation~\ref{eq:moisture}, the condensation term $C$ which drives precipitation is itself a functional of the prognostic state (humidity, temperature, pressure, and winds). Therefore, re-predicting precipitation at each step solely from the current prognostic state is physically reasonable, and its temporal continuity across steps is inherited from the evolving prognostic state rather than from any autonomous dynamics of precipitation itself.

Interestingly, precipitation admits an alternative modeling paradigm known as \textit{nowcasting}, which falls under the evolution operator learning framework (Equation~\ref{eq:autoregressive-formula}). In contrast to the diagnostic treatment described above where each timestep's estimate is fully independent of previous precipitation values, nowcasting treats precipitation as a variable that evolves under its own past values, without explicit conditioning on external physical drivers~\citep{shi2015convolutional, shi2017deep, ravuri2021skilful}. This approach is supported by two key arguments:

\begin{enumerate}
  \item Complexity of process variability: The underlying dynamics are often intricate and stochastic in nature, making explicit driver-based modeling difficult 
  to formulate and generalize.
  \item Implicit driver signatures: Since the physical drivers are themselves autocorrelated in time, their influence leaves detectable imprints on the target variable's historical record, allowing past values to serve as indirect proxies for those drivers.
\end{enumerate}

Beyond precipitation nowcasting, this rationale is similarly well-grounded in hydrological time series forecasting, particularly for streamflow. The effects of antecedent precipitation, soil moisture, and temperature are embedded in the runoff record, making the variable's own history an informative, if indirect, reflection of its physical forcings~\citep{labat2005recent, shiri2010short, zhang2018univariate}.

\subsection{Atmospheric Tracer Transport}
A second, richer example comes from atmospheric tracer transport. What makes it particularly instructive is that it can be viewed at two different levels of abstraction—each revealing a different modeling challenge. At a finer level, the tracer transport equation itself contains a physical process that must be parameterized: vertical convection. At a coarser level, the entire tracer transport model can be viewed as a parameterized physical process coupled within a larger NWP model. Both perspectives are worth examining.

When numerically integrating the continuity equation for a tracer $\mu$ over a discrete grid, the evolution is typically split into horizontal advection, vertical convection, and sink/source terms as follows: 

\begin{equation}
\frac{\partial\mu}{\partial t} + u\frac{\partial\mu}{\partial x} + v\frac{\partial\mu}{\partial y} + w\frac{\partial\mu}{\partial z} = \Sigma
\label{eq:transport-formula}
\end{equation}

Here, the tracer concentration $\mu$ is the prognostic variable. It carries an explicit time derivative and is stepped forward in time. The vertical velocity ($w$), however, is a diagnostic variable that drives the vertical convection (the fourth term on the left-hand side of Equation~\ref{eq:transport-formula}). It has no evolution equation of its own and is instead determined instantaneously from the current atmospheric state, such as temperature ($T$), specific humidity ($q$), geopotential height ($z$), and other variables depending on the specific parameterization scheme. Therefore, learning a surrogate for $w$ is a state-conditioned operator evaluation task (Equation~\ref{eq:standard-regression}):

\begin{equation}
w = \mathcal{G}_{\theta}(T, q, z, \dots)
\end{equation}

Before moving to the coarser level, it is instructive to rearrange the tracer transport equation to better align with the ML formalizations discussed earlier. If, instead of parameterizing only the vertical convection component, one wishes to use a neural network to surrogate all advection and source/sink dynamics simultaneously, the equation can be recast to isolate the local time derivative. This frames the problem in terms of the overall instantaneous tendency of the tracer~\citep{benson2025atmospheric}:

\begin{equation}
\frac{\partial \mu}{\partial t} = \mathcal{N}_{\theta}(\mu,\, u,\, v,\, w,\, T,\, q,\, z,\, \dots)
\end{equation}

\noindent
Under this scientific formulation, the neural network acts as a complete \textbf{tendency operator}. It learns the instantaneous rate of change for the prognostic variable $\mu$ based entirely on the concurrent atmospheric state. Once the operator $\mathcal{N}_{\theta}$ produces this approximate tendency, the prognostic tracer is marched forward using a standard numerical integration scheme:

\begin{equation}
\mu_{t+1} = \mu_t + \left( \frac{\partial \mu}{\partial t} \right) \Delta t
\end{equation}

\noindent
This procedure naturally maps to an \textbf{autoregressive rollout}: the learned tendency operator is applied repeatedly, step-by-step, to evolve the prognostic state forward in time (Equation~\ref{eq:autoregressive-tendency-formula}).

However, a data-driven perspective motivates a completely different framing: the spatiotemporal predictive learning approach~\citep{erfani2025interactive, erfani2026spatiotemporal}. Given a sequence $\mathbf{X}^{t,T} = \{x_{t-T+1}, \dots, x_t\}$ representing the past $T$ frames up to time $t$, model aims to predict the subsequent $T'$ frames $\mathbf{Y}^{t+1,T'} = \{x_{t+1}, \dots, x_{t+T'}\}$. Here, each frame $x_i \in \mathbb{R}^{C \times H \times W}$ is treated as an image with $C$ channels, height $H$, and width $W$. In practice, these sequences are represented as high-dimensional tensors, i.e., $\mathbf{X} \in \mathbb{R}^{T \times C \times H \times W}$. A model with learnable parameters $\theta$ then learns a mapping $\mathcal{N}_{\theta} : \mathbf{X}^{t,T} \mapsto \mathbf{Y}^{t+1,T'}$ by extracting combined spatial and temporal dependencies (Equaiton~\ref{eq:seq2seq}).

This represents a fundamental shift in perspective. Under the scientific formulation, the neural network approximates the tracer tendency, which is then rolled forward using explicit numerical integration. In this view, the tracer remains a strict prognostic variable governed by evolution operator learning. In the free-form data-driven approach, however, the physical equations are set aside. When the entire transport problem is reduced to multichannel spatiotemporal arrays, the surrogate is no longer an explicit tendency operator. Instead, it becomes a spatiotemporal sequence model, often realized through a recurrent architecture, that maps the recent history of atmospheric states directly to the tracer concentration field at the next time step.

\section{Conclusion}
In this work, we distinguish between two fundamentally different approaches to applying ML in atmospheric modeling. While both aim to replace computationally expensive functions with efficient approximations, they diverge entirely in their underlying philosophy.

The first approach is embedded surrogate modeling. Here, ML is integrated inside an existing governing system. The goal is physics-aware substitution, i.e., replacing a specific component (such as a diagnostic closure or physical tendency) while strictly preserving the overall structure, constraints, and ontology of the physical model. The ultimate aim is computational efficiency without breaking the established physical mechanisms. The second approach is free-form data-driven modeling. Here, the original physical system is treated as an unknown function. The model makes no commitment to governing equations and internal mechanisms. Instead, the task is defined entirely by the statistical and geometrical structure of the data, representing data by their (spatial or temporal) dependencies, where the goal is to learn latent patterns rather than the system itself.

Through a historical survey and mathematical formalization of MLWP and its components, this work traces how the field initially developed and how it has evolved to its current state. Moving forward, choosing the appropriate ML architecture requires a deliberate and systematic approach that consciously aligns model design with either the physical problem structure or the data characteristics. Both traditions offer immense value, but they cannot be evaluated by the same standard. Embedded surrogates must be judged by whether they preserve the physical meaning and stability of the governing system. Free-form spatiotemporal models, conversely, must be judged by how effectively they forecast future states from the data representations they are given.

\bibliographystyle{plainnat}
\bibliography{ref.bib}

\newpage
\appendix
\section{ML Architecture Recommendations via AI Chatbot}
\label{sec:appendix-a}

The following prompt was designed and presented to several AI chatbots to elicit model recommendations. The prompt deliberately strips atmospheric data of its physical semantics, representing the forecasting problem purely as a high-dimensional spatiotemporal tensor prediction task.

\begin{tcolorbox}[colback=gray!10, colframe=gray!50, sharp corners]
Based on the following description of my data, what type of ML model would 
you recommend? Do not consider what we have already discussed in our previous chats.

\medskip
Consider a high-dimensional, regularly gridded array evolving 
over discrete time steps. At each time $t$, the system state is represented 
as a multi-channel tensor
\[
  X_t \in \mathbb{R}^{C \times H \times W},
\]
where $C$ denotes the number of channels (features), and $H \times W$ defines 
a fixed 2D lattice. 

\medskip
Given an arbitrary sequence of past states $\{X_{t-k}, \dots, X_t\}$, 
the objective is to learn a mapping that predicts one or more future states 
$\{X_{t+1}, \dots, X_{t+m}\}$. The data exhibit strong spatiotemporal dependencies, including:

\begin{itemize}
  \item Local correlations across neighboring grid points
  \item Nonlocal interactions across distant regions
  \item Temporal continuity with both short-term and long-range dependencies
  \item Multi-scale structure, where patterns evolve at different spatial 
        and temporal resolutions
\end{itemize}

\medskip
The learning task can be formulated either as:
\begin{itemize}
  \item \textbf{Direct state prediction:} learning $X_{t+1} = f(X_t)$, or
  \item \textbf{Increment prediction:} learning $\Delta X_t = g(X_t)$ such 
        that $X_{t+1} = X_t + \Delta X_t$
\end{itemize}

\medskip
No assumptions are made about the underlying generative process, the semantics 
of the channels, or any governing equations. The problem is treated purely as 
learning a mapping between sequences of high-dimensional tensors.
\end{tcolorbox}

The AI chatbot recommendations are summarized in Table~\ref{tab:model_comparison}. Across all five systems, spatiotemporal model families received the highest overall priority. Within this category, however, there is a notable split: four out of five models (Gemma, Gemini, Claude, and ChatGPT) ranked \textbf{Transformer-based} architectures first, citing their ability to capture long-range spatial and temporal dependencies through self-attention mechanisms. Perplexity was the lone exception, assigning first priority to \textbf{recurrent} models (e.g., ConvLSTM), likely emphasizing temporal continuity as the primary structural constraint.

Neural Operators, specifically the Fourier Neural Operator (FNO), emerged as the second-ranked recommendation across four of the five chatbots. This consensus reflects the explicit mention of nonlocal spatial interactions in the prompt: operating in the frequency domain, a single FNO layer can correlate information across the entire spatial grid instantaneously, making it particularly well-suited to problems with global dependencies. ChatGPT recommended a broader set of hybrid architectures including a CNN encoder paired with a temporal Transformer, and a CNN combined with a state-space model (Mamba), reflecting a stronger emphasis on modular design. ConvLSTM appeared only as a fifth-priority option in ChatGPT's ranking.

Taken together, the LLM responses converge on a two-tier recommendation: \textbf{spatiotemporal Transformers} as the primary architecture for capturing multi-scale dependencies, and \textbf{neural operators} (particularly FNO) as a powerful complement for handling nonlocal spatial structure. This convergence provides independent, data-agnostic motivation for the model families explored in this work.

\begin{table}[h]
    \centering
    \caption{ML model architectures recommended by five AI chatbots.}
    \label{tab:model_comparison}
    \begin{tabular}{lcccc}
        \toprule
        \textbf{Chatbot} & \multicolumn{3}{c}{\textbf{Spatiotemporal Model}} & \textbf{Neural Operators} \\
        \cmidrule(lr){2-4}
        & Transformers & Recurrent & 3D Convolution & \\
        \midrule
        Gemma      & 1 & 3 & —  & 2 \\
        Gemini     & 1 & 3 & 4  & 2 \\
        Claude     & 1 & 3 & 3  & 2 \\
        ChatGPT\textsuperscript{\dag} & 1 & 5 & — & 3 \\
        Perplexity & 3 & 1 & 2  & — \\
        \bottomrule
    \end{tabular}
    \vspace{4pt}
    \begin{minipage}{\linewidth}
        \footnotesize
        \textsuperscript{\dag} ChatGPT additionally recommended a ``CNN encoder + 
        temporal Transformer'' (2nd priority) and ``CNN + state-space model,'' 
        e.g., Mamba (4th priority). These hybrid categories were omitted from 
        the table as no other chatbot proposed equivalent architectures. 
        ConvLSTM was ranked 5th by ChatGPT and is captured under the 
        Recurrent column.
    \end{minipage}
\end{table}

\end{document}